%% file: main.tex
\documentclass[conference, a4paper]{IEEEtran}
\IEEEoverridecommandlockouts
\usepackage{amsmath,amssymb,mathtools,amsthm,bm}
 \usepackage{cite}
\usepackage{url}
\usepackage{physics}
\usepackage{graphicx}
\usepackage{bbm}
\usepackage{dsfont}
\usepackage{enumitem}
\usepackage{lipsum}
\usepackage{booktabs}
\usepackage{comment}
\usepackage[caption=false,font=footnotesize]{subfig}
\usepackage[hidelinks]{hyperref}
\usepackage{doi}
\usepackage{flushend}
\usepackage{algorithmic}
\usepackage{algorithm}
\usepackage{soul}
\usepackage{tikz}
\usepackage{flushend}
\usepackage{multirow}

\usetikzlibrary{shapes.geometric, arrows.meta, positioning, calc}
\usepackage{pgfplots}
\pgfplotsset{compat=1.18}   
\usepackage[caption=false,font=footnotesize]{subfig}

\usepackage[T1]{fontenc}
\usepackage{lmodern}

\usepackage{times}

\newif\ifshowchanges
\showchangesfalse   

\definecolor{revisionGB}{HTML}{0568A7}
\DeclareRobustCommand{\rev}[1]{%
  \begingroup
  \ifshowchanges
    \color{revisionGB}%
  \fi
  #1%
  \endgroup
}

\begin{document}
%
\title{Modal Analysis of Spatial Load Correlation in AI Data Center-Dominated Power Systems \vspace{-0.5em}}

\author{
\IEEEauthorblockN{
Chandan Chaudhary\IEEEauthorrefmark{1}, 
Michael Murillo\IEEEauthorrefmark{2}, 
Mohammed Ben-Idris\IEEEauthorrefmark{1}, 
Joydeep Mitra\IEEEauthorrefmark{1}, 
Dilip Pandit\IEEEauthorrefmark{3}, 
Atri Bera\IEEEauthorrefmark{3}
}
\IEEEauthorblockA{\IEEEauthorrefmark{1}
Electrical and Computer Engineering, Michigan State University, East Lansing, MI 48824, USA
}
\IEEEauthorblockA{\IEEEauthorrefmark{2}
Computational Mathematics, Science and Engineering, Michigan State University, East Lansing, MI 48824, USA
}
\IEEEauthorblockA{\IEEEauthorrefmark{3}
Sandia National Laboratories, Albuquerque, NM 87185, USA
}
\IEEEauthorblockA{
Emails: \{chaud152, murillom, benidris, mitraj\}@msu.edu, \{dpandit, abera\}@sandia.gov \vspace{-1.5em}
}
}

{\maketitle}

\begin{abstract}
\input{Contents/Abstract}
\end{abstract}


%
\IEEEpeerreviewmaketitle

\vspace{-0.75em}
\section{Introduction}
\vspace{-0.25em}
\input{Contents/Introduction}

\vspace{-0.5em}
\section{Spatial Load Correlation}
\label{sec:spatialloadcorr}
\input{Contents/Spatial_load_correlation}

\vspace{-0.5em}
\section{Methodology}
\label{sec:dmd}
\enlargethispage{\baselineskip}
\input{Contents/DMD_framework}

\vspace{-1em}
\section{Case Study and Analysis}
\enlargethispage{\baselineskip}
\label{sec:casestudyandresults}
\input{Contents/Sim_Study}
\input{Contents/Results}

\section{Conclusion}
\label{sec:conclusion}\enlargethispage{\baselineskip}
\input{Contents/conclusion}

\section*{Acknowledgment}
\input{Contents/Acknowledgment}

\enlargethispage{6\baselineskip}
\vspace{-0.5em}
\bibliography{references}
\bibliographystyle{ieeetr}

\end{document}

%% file: Contents/Abstract.tex
Hyperscale AI data centers induce spatially and temporally correlated load fluctuations that violate classical independence assumptions and are not captured by time-averaged spectral methods. These correlations are episodic and non-stationary, so they demand analysis that resolves transient structure. This paper applies Dynamic Mode Decomposition (DMD) to the temporal evolution of pairwise inter-bus correlation coefficients and forms a low-dimensional state representation that enables modal analysis without a stationarity assumption. The recovered modes distinguish sustained coherence, decaying transients, and intensifying events, and their oscillation timescales map to underlying physical coupling mechanisms. The method is evaluated on an IEEE 39-bus Real-Time Digital Simulator (RTDS) testbed with three converter-interfaced AI data center loads driven by synthetic workload profiles. \rev{A global analysis attributes the dominant correlation energy to a slow thermal band, and a sliding-window analysis identifies brief intensification events in a small fraction of windows that align with stochastic workload coincidences. Cross-validation with RTDS voltage coherence confirms elevated coupling during these intervals.} The proposed modal growth indicator provides an early-warning signal of correlation intensification, with a lead of of about 4~s before pairwise coherence reaches its peak.

%% file: Contents/Introduction.tex
\enlargethispage{5\baselineskip}
The electric grid is undergoing a structural transformation driven by hyperscale AI computing infrastructure. Data centers accounted for approximately 4.4\% of total U.S. electricity consumption in 2023, with projections rising to 6.7--12\% by 2028~\cite{shehabi2024usdc}. Of the 166~GW of forecast peak load growth in the United States through 2030, roughly 90~GW is attributed to data-center expansion~\cite{nerc2025largeloads}. 
Unlike conventional loads, converter-interfaced AI data centers provide negligible inertia and impose rapid, quasi-periodic transients from mini-batch gradient synchronization and thermal cycling. Recent surveys and field studies identify these characteristics as emergent grid risks~\cite{xiong2020modeling , Cheng2026AIGrid, seshmasetti2025review}. Field observations document 14.7~Hz oscillations from converter control interactions with network impedances ~\cite{mishra2025understanding}, and large-scale AI workloads have been shown to excite wide-area oscillatory modes overlapping weakly damped inter-area frequencies~\cite{ko2026wide}. Operational risk assessments on benchmark transmission models confirm that the fast ramp patterns and demand fluctuations characteristic of large data center clusters can drive both transient and small-signal instabilities across regional networks~\cite{kwon2025operational,paccou2025exploring}.

Beyond individual-facility effects, multiple physical mechanisms drive \textit{spatially synchronized} operation across geographically distributed data center clusters \cite{Chaudhary2026SpatialAI}. Distributed deep learning training flushes gradient buffers at every mini-batch boundary and produces quasi-periodic coordinated power swings at 2--30s intervals across all GPU nodes~\cite{Chaudhary2025DataCenterStability, go2025characterizing}. Shared thermal management under common environmental conditions aligns Heating Ventilation and Air Conditioning (HVAC) cycles across co-located facilities~\cite{Chaudhary2026SpatialAI}, and shared network impedances couple the converter control loops of electrically proximate facilities. Together, these mechanisms drive inter-bus power fluctuations that alternate between near-perfect synchrony and anti-correlation within tens of seconds as training jobs transition between computational and communication phases~\cite{li2024unseen, Chaudhary2026SpatialAI}. These fluctuations produce a non-stationary inter-bus coherence structure that intensifies and dissolves episodically.

\rev{The spatial correlation of load dynamics across distributed AI data centers remains insufficiently characterized in the power systems literature. Our prior work establishes that this correlated behavior exists across geographically distributed facilities, identifies its three physical mechanisms~\cite{Chaudhary2026SpatialAI}, and quantifies the resource adequacy risk of correlated large loads~\cite{chaudhary2026adequacy}}. Recent studies \cite{Cheng2026AIGrid, seshmasetti2025review, kwon2025operational} examine instability at individual facilities and leave inter-bus coherence unaddressed. Current grid planning and monitoring frameworks still treat loads at different buses as independent. System operators lack tools to detect, quantify, and track spatially correlated load behavior across sites. Welch's cross-spectral periodogram \cite{bendat2011random} assumes stationarity and cannot resolve episodic coherence. Prony decomposition \cite{hauer2002application} targets quasi-stationary electromechanical oscillations and misses the spatial dependence structure of converter-dominated load clusters. Three gaps follow. No method tracks inter-bus correlation in operation. Planning frameworks assume independence, and standard spectral and modal tools assume stationarity.

\rev{This paper proposes a data-driven modal framework based on Dynamic Mode Decomposition (DMD) \cite{vicario2023dmd}. The framework makes three contributions. The first applies DMD to the temporal evolution of the inter-bus correlation state vector rather than to raw power, which exposes coherence dynamics without a stationarity assumption. The second introduces a modal growth indicator that flags correlation intensification before the pairwise coefficients reach their peak. The third cross-validates the indicator against RTDS voltage coherence on an IEEE 39-bus testbed. These contributions turn the existence result of \cite{Chaudhary2026SpatialAI} into an operational detection method.}
\enlargethispage{5\baselineskip}

The remainder of this paper is organized as follows. Section~\ref{sec:spatialloadcorr} reviews the concept of spatial load correlation from \cite{Chaudhary2026SpatialAI}. Section~\ref{sec:dmd} presents the proposed DMD-based model evaluation. Section~\ref{sec:casestudyandresults} describes the case study and analyzes the correlation dynamics and modal behavior. Section~\ref{sec:conclusion} concludes the paper and outlines directions for future research.

%% file: Contents/Spatial_load_correlation.tex
Spatial load correlation refers to the statistical dependence among active power fluctuations at different transmission buses. Under classical planning assumptions, loads at geographically separated buses vary independently so that aggregate fluctuations scale with the square root of the number of demand elements \cite{Chaudhary2026SpatialAI}. When loads are correlated, aggregate fluctuations scale faster, diversity factors understate variance, and reserve margins sized for independence are insufficient \cite{chaudhary2026adequacy}. The characterization of this phenomenon in AI data center-dominated grids, the physical mechanisms that produce inter-bus coherence, and their implications for voltage and frequency stability are established in prior work~\cite{Chaudhary2026SpatialAI}. 

%% file: Contents/DMD_framework.tex
The proposed method constructs a correlation state-space from raw RTDS recordings and, through spectral characterization and DMD, extracts dominant spatial correlation modes and their physical interpretations.

\enlargethispage{4\baselineskip}
\subsection{Correlation State-Space Formulation}
Let $\Delta P_i(t)$ denote the zero-mean active power fluctuation at bus~$i$, with the slow trend removed by a 30\,s moving-average filter. For $N_b$ buses with data center loads, the pairwise Pearson correlation over a window of length $T_w$ is \cite{birihanu2025explainable}:
\begin{equation}
\rho_{ij}(t) = \frac{\displaystyle\sum_{s=0}^{T_w-1} \Delta P_i(t{-}s)\,\Delta P_j(t{-}s)}{\sqrt{\displaystyle\sum_s \Delta P_i^2(t{-}s) \cdot \sum_s \Delta P_j^2(t{-}s)}}
\label{eq:sliding_corr}
\end{equation}
The $N_p = N_b(N_b-1)/2$ unique pairwise coefficients are assembled into a state vector:
\begin{equation}
\mathbf{x}(t) = \bigl[\rho_{12}(t),\;\rho_{13}(t),\;\ldots,\;\rho_{(N_b-1)N_b}(t)\bigr]^\top \in \mathbb{R}^{N_p}
\label{eq:state_vector}
\end{equation}
This state vector isolates the spatial coherence structure from individual load and single-bus ramp transients, and is the natural choice for modal analysis of aggregate fluctuation statistics and correlated contingency exposure. \rev{Because the Pearson coefficient is amplitude-normalized, the state reports the directional alignment between buses and not their fluctuation magnitude, so two facilities that ramp hard but independently register near zero rather than the large value a raw-power covariance would assign.} Equivalent bus voltage magnitudes serve as the network-level validation channel.

\subsection{Spectral Concentration and Limitations of Time-Averaging}
The spatial covariance matrix $\mathbf{R} = \mathbb{E}[\Delta\mathbf{P}\,\Delta\mathbf{P}^\top]$ and its eigendecomposition $\mathbf{R} = \sum_k \lambda_k \mathbf{u}_k \mathbf{u}_k^\top$ provide a global summary of coherence structure. The spectral concentration index:
\begin{equation}
\rho_s = \frac{\lambda_1}{\sum_{k=1}^{N_b} \lambda_k}
\label{eq:corr_index}
\end{equation}
measures the fraction of total load variance attributable to the dominant spatial mode. For $N_b$ independent loads, $\rho_s \approx 1/N_b$. A strongly correlated ensemble drives $\rho_s$ toward unity. The magnitude-squared coherence:
\begin{equation}
\gamma_{ij}^2(\omega) = \frac{|S_{ij}(\omega)|^2}{S_{ii}(\omega)\,S_{jj}(\omega)}
\label{eq:coherence}
\end{equation}
extends this analysis to the frequency domain~\cite{bendat2011random}. Both $\rho_s$ and $\gamma_{ij}^2(\omega)$ are expectations over an observation window, so a strongly coherent episode and a decorrelated episode yield an intermediate average that accurately describes neither state. Frequency resolution scales as $1/T_w$, so a window sufficient for spectral convergence is long enough to obscure episode structure. This motivates a method that operates on the time evolution of the state vector directly, without a stationarity requirement.

\subsection{DMD Algorithm}
Given the correlation state vector sequence at uniform time steps $\Delta t$, the snapshot matrices are:
\begin{align}
\mathbf{X}  &= \bigl[\mathbf{x}_1,\;\ldots,\;\mathbf{x}_{m-1}\bigr] \in \mathbb{R}^{N_p \times (m-1)} \label{eq:X_mat} \\
\mathbf{X}' &= \bigl[\mathbf{x}_2,\;\ldots,\;\mathbf{x}_m\bigr] \in \mathbb{R}^{N_p \times (m-1)} \label{eq:Xp_mat}
\end{align}
DMD seeks the best-fit linear operator $\mathbf{A}$ such that $\mathbf{X}' \approx \mathbf{A}\mathbf{X}$~\cite{schmid2010dynamic, kutz2016dynamic}. Direct computation of $\mathbf{A} \in \mathbb{R}^{N_p \times N_p}$ is avoided to favor the low-rank projection from the economy SVD of $\mathbf{X}$:
\begin{equation}
\vspace{-0.5em}
\mathbf{X} = \mathbf{U}\,\boldsymbol{\Sigma}\,\mathbf{V}^\top, \quad \mathbf{U} \in \mathbb{R}^{N_p \times r}
\label{eq:svd}
\end{equation}
where the truncation rank $r$ is determined by a gap criterion on the singular value spectrum. The smallest index at which $\sigma_k / \sigma_{k+1}$ exceeds 1.3 is taken as $r$. When no such gap exists, a fixed fallback rank is used and the flat spectrum is recorded as evidence of non-stationary dynamics. The reduced operator projected onto the dominant Proper Orthogonal Decomposition (POD) subspace is:
\begin{equation}
\widetilde{\mathbf{A}} = \mathbf{U}^\top \mathbf{X}' \mathbf{V}\,\boldsymbol{\Sigma}^{-1} \in \mathbb{R}^{r \times r}
\label{eq:reduced_A}
\end{equation}
Eigendecomposition $\widetilde{\mathbf{A}}\mathbf{W} = \mathbf{W}\boldsymbol{\Lambda}$ yields eigenvalues $\{\mu_k\}$ and eigenvectors $\{\mathbf{w}_k\}$. The full-dimensional DMD modes are:
\begin{equation}
\boldsymbol{\phi}_k = \mathbf{X}'\,\mathbf{V}\,\boldsymbol{\Sigma}^{-1}\mathbf{w}_k \in \mathbb{R}^{N_p}
\label{eq:dmd_modes}
\end{equation}
Mode amplitudes $b_k$ are obtained by least-squares projection of the first snapshot onto the mode basis.
\enlargethispage{4\baselineskip}

\subsection{Physical Interpretation of Eigenvalues}
The eigenvalues $\{\mu_k\}$ are the primary diagnostic output. Two continuous-time quantities are recovered from each discrete-time eigenvalue:
\begin{align}
f_k      &= \frac{\mathrm{Im}(\ln \mu_k)}{2\pi\,\Delta t} \quad (\text{oscillation frequency, Hz}) \label{eq:freq} \\
\sigma_k &= \frac{\mathrm{Re}(\ln \mu_k)}{\Delta t} \quad (\text{growth/decay rate, s}^{-1}) \label{eq:growth}
\end{align}
The position of $\mu_k$ on the complex plane carries direct physical meaning. An eigenvalue on the unit circle ($|\mu_k|=1$, $\sigma_k=0$) corresponds to sustained oscillatory coherence driven by an active periodic mechanism such as settled HVAC cycling. An eigenvalue strictly inside the unit circle ($|\mu_k|<1$, $\sigma_k<0$) indicates a coherence burst in natural decay. An eigenvalue outside the unit circle ($|\mu_k|>1$, $\sigma_k>0$) signals intensifying coherence. The positive growth rate is detectable before the pairwise coefficients $\rho_{ij}$ reach their peak, and the time to peak is approximated by $1/\sigma_k$. The frequency $f_k$ maps to the physical coupling mechanism via Table~\ref{tab:mode_map}.

\subsection{Mode Energy, Shape, and Reconstruction Fidelity}
The mode amplitudes $b_k$ are obtained by least-squares projection of the first snapshot onto the mode basis~\cite{Tu2014OnDMD}. The energy contribution of each mode is~\cite{kutz2016dynamic}:
\begin{equation}
E_k = \frac{|b_k|^2 \,\|\boldsymbol{\phi}_k\|^2}{\sum_{j=1}^r |b_j|^2\,\|\boldsymbol{\phi}_j\|^2}
\label{eq:mode_energy}
\end{equation}
$E_k$ is the correct measure of instantaneous dominance and is not equivalent to the eigenvalue magnitude rank. State reconstruction at time $t$ follows from $\mathbf{x}(t) \approx \sum_k b_k \boldsymbol{\phi}_k \mu_k^{t/\Delta t}$~\cite{Tu2014OnDMD}, and the rank-$r$ reconstruction error
confirms whether a compact modal description is adequate~\cite{kutz2016dynamic}. Each mode $\boldsymbol{\phi}_k \in \mathbb{R}^{N_p}$ has one entry per bus pair. A mode at the workload orchestration frequency with roughly uniform entries across all pairs indicates a centralized scheduler that acts simultaneously on all facilities. A mode at converter frequencies with entries concentrated on electrically proximate pairs reflects impedance-mediated coupling. The joint use of $f_k$ and $\boldsymbol{\phi}_k$ provides two independent channels for physical mechanism attribution.
\enlargethispage{5\baselineskip}

\subsection{Sliding-Window Portrait}
A single global DMD characterizes the average modal structure over the full record and is not appropriate for non-stationary dynamics~\cite{Brunton_Kutz_2022}. The proposed approach applies DMD over windows of length $T_w^{\mathrm{DMD}}$ advanced in steps of $\delta t$. Each window position $n$ yields the triple $\{(\mu_k^{(n)}, \boldsymbol{\phi}_k^{(n)}, E_k^{(n)})\}$. The dominant mode frequency and energy plotted against window index constitute a time-frequency portrait of the correlation dynamics. Unlike a short-time Fourier spectrogram, which populates all frequency bins regardless of data content, the DMD portrait assigns energy only to modes the data actively supports within each window~\cite{Brunton_Kutz_2022}, and is therefore a more physically interpretable operational diagnostic.

\subsection{Analysis Method}
The complete analysis pipeline is summarized in Algorithm~\ref{alg:full_protocol}. 
Raw active power recordings are detrended and assembled into the correlation state vector $\mathbf{x}(t)$ via~\eqref{eq:state_vector}, with window $T_w = 10$\,s and step 1\,s. \rev{The window length is chosen to span at least one full cycle of the slowest mechanism it must resolve while remaining short enough to track episode boundaries.} Global DMD is then applied to the full record: the singular value gap determines the truncation rank, each eigenvalue $\mu_k$ is mapped to a physical mechanism via Table~\ref{tab:mode_map}, and reconstruction error confirms whether the rank-$r$ global description is adequate. The sliding-window portrait follows, with $m \geq 3N_p$ snapshots per window as the minimum condition for SVD conditioning, and windows where $|\mu_k^{(n)}| > 1$ are flagged as high-correlation events.

\enlargethispage{\baselineskip}

\begin{table}[!ht]
\centering
\vspace{-1em}
\caption{DMD Mode Frequency Bands and Physical Mechanism Correspondence}
\label{tab:mode_map}
\vspace{-0.5em}
\begin{tabular}{lll}
\toprule
\textbf{Freq.\ Band } & \textbf{Physical Mechanism} & \textbf{Timescale} \\
\midrule
0.003--0.134\,Hz & Slow / Thermal (HVAC, chiller drift) & Minutes \\
0.134--1.0\,Hz   & Workload orchestration (GPU mini-batch) & Seconds \\
1.0--26.1\,Hz    & Converter control coupling           & Sub-second \\
\bottomrule
\end{tabular}
\vspace{-1em}
\end{table}
\rev{The band boundaries in \ref{tab:mode_map} are data-driven rather than assumed. The 0.134\,Hz edge is the power spectral density transition above which mini-batch orchestration energy gives way to faster workload-switching variance, the 1.0\,Hz edge coincides with the converter outer-loop control bandwidth, and the 26.1\,Hz upper edge is the frequency at which inter-bus voltage coherence falls below 0.3. The boundaries mark dominant mechanisms rather than hard partitions, so adjacent mechanisms overlap near each edge. The 2--30\,s mini-batch interval, near 0.17--0.5\,Hz, falls inside the workload band.}

The cross-validation step computes $\gamma^2_{ij}(\omega)$ of the per-bus RTDS voltage deviations $\Delta V_i(t)$ at flagged and sparse portrait episodes. System frequency $f(t)$ in a synchronous network is a single system-wide scalar and carries no inter-bus spatial information; $\Delta V_i(t)$ is the correct network observable because it is spatially differentiated across buses and its inter-bus coherence directly reflects coupling between data center buses. Elevated $\gamma^2_{ij}(\omega)$ at flagged DMD episodes confirms that the modal portrait reflects genuine network-level coupling rather than a state vector construction artifact. 

\begin{algorithm}[!ht]
\caption{DMD-Based Spatial Correlation Analysis}
\label{alg:full_protocol}
\begin{algorithmic}[1]
\REQUIRE AI load profiles $\{P_i(t), Q_i(t)\}$; RTDS recordings $\{\Delta P_i,\,V_i\}$ at 60\,Hz (per-bus); system frequency $f(t)$ logged as a scalar reference
    \STATE Detrend: $\Delta P_i(t)\! \leftarrow\! P_i(t)\!-\!\bar{P}_i(t)$ via 30s moving average
\STATE Compute $\rho_{ij}(t)$ via~\eqref{eq:sliding_corr}; assemble $\mathbf{x}(t)$ via~\eqref{eq:state_vector}
\STATE Form $\mathbf{X}$, $\mathbf{X}'$; compute SVD~\eqref{eq:svd}; select rank $r$ from singular value gap
\STATE Compute $\widetilde{\mathbf{A}}$~\eqref{eq:reduced_A}, $\{\mu_k\}$, $\{\boldsymbol{\phi}_k\}$~\eqref{eq:dmd_modes}; recover $f_k$, $\sigma_k$
\FOR{each window position $n$ with length $T_w^{\mathrm{DMD}}$, step $\delta t$}
  \STATE Repeat steps 3--4; record $f_k^{(n)}$, $E_k^{(n)}$~\eqref{eq:mode_energy}, $|\mu_k^{(n)}|$; flag if $|\mu_k^{(n)}| > 1$
\ENDFOR
\STATE Cross-validate: compute $\gamma^2_{ij}(\omega)$ of $\Delta V_i(t)$ at flagged episodes vs.\ sparse episodes
\end{algorithmic}
\end{algorithm}

\enlargethispage{5\baselineskip}

%% file: Contents/Sim_Study.tex
\begin{figure*}[!htbp]
  \centering

  \begin{minipage}[t]{0.39\linewidth}
    \centering
    \includegraphics[width=\linewidth]{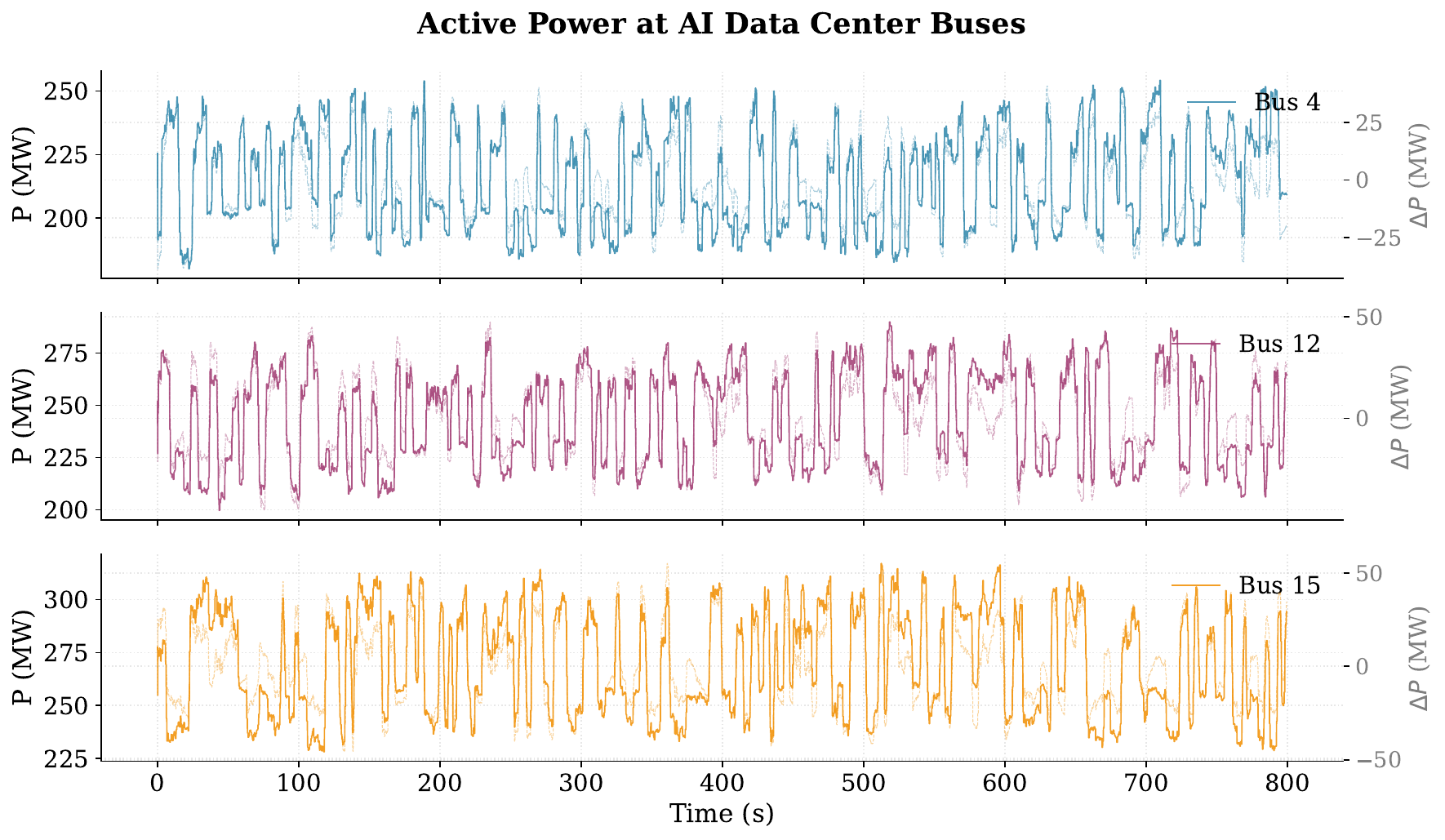}
  \end{minipage}
  \hfill
  \begin{minipage}[t]{0.60\linewidth}
    \centering
    \includegraphics[width=\linewidth]{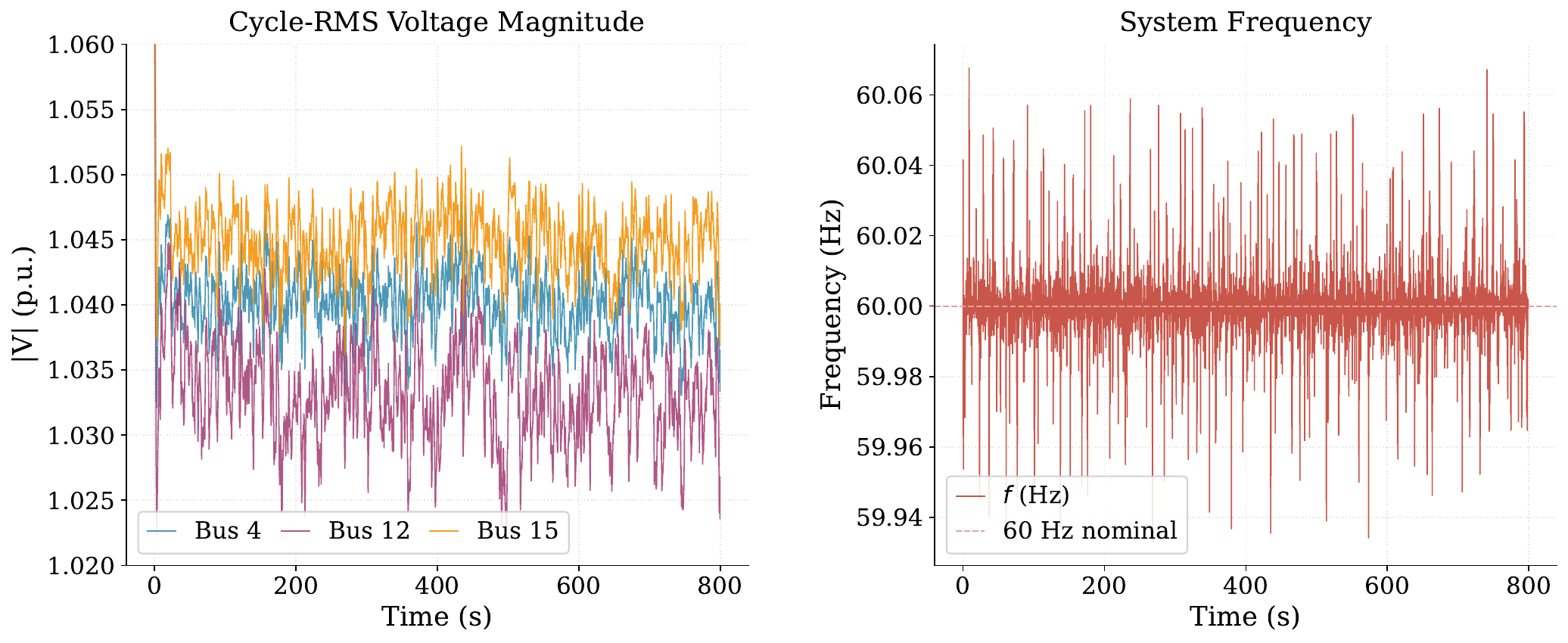}
  \end{minipage}

  \vspace{-1em}
  \caption{Left: Timeseries load profile of Synthetic Data Center Load. Right: bus voltage magnitudes (p.u.) at buses 4, 12, and 15, and system frequency deviation.}
  \vspace{-1em}
  \label{fig:time_series_load_volt_freq}
\end{figure*}

This section presents the AI data center load model and the test system, then analyzes the resulting correlation dynamics and modal behavior to validate the proposed methodology.

\subsection{\rev{Synthetic AI Data Center Load Profile Model}}
Real-time transmission-level measurements of AI data center loads are not publicly available, and RTDS validation requires a high fidelity input signal with physically realistic spectral and statistical properties. A component-wise stochastic model is therefore constructed for each facility, calibrated to published GPU cluster power characterizations~\cite{Chaudhary2025DataCenterStability, go2025characterizing, li2024unseen}.
The aggregate active power is:
\begin{equation}
P_i(t) = P_{\mathrm{IT},i}(t) + P_{\mathrm{HVAC},i}(t) + P_{\mathrm{Ch},i}(t) + P_{\mathrm{Aux},i}(t)
\label{eq:aggregate}
\end{equation}
The IT layer is the primary driver of inter-bus correlation and is modeled as a semi-Markov process with three states that correspond to the dominant phases of a distributed training iteration: Surge (GPU forward and backward propagation), Dip (AllReduce gradient synchronization barrier), and Base (idle or data-loading) \cite{Chaudhary2025DataCenterStability, go2025characterizing}. State sojourn times are drawn from $\mathcal{U}(2,30)$\,s, consistent with reported mini-batch periods for large-model training~\cite{go2025characterizing}. Transitions are ramp-rate limited at $R_i \sim \mathcal{U}(25,60)$\,MW/s. An Ornstein--Uhlenbeck (OU) process overlays state-dependent noise to represent within-state GPU variability \cite{bibbona2008ornstein, go2025characterizing}:
\begin{equation}
d\xi_{\mathrm{IT}} = -\theta_{\mathrm{IT}}\,\xi_{\mathrm{IT}}\,dt + \sigma_{\mathrm{IT}}(s)\,dW_t, \quad \theta_{\mathrm{IT}} = 0.8\,\mathrm{s}^{-1}
\label{eq:ou_it}
\end{equation}
HVAC and Chiller loads track IT load through first-order thermal lags with time constants $\tau_{\mathrm{HVAC}} = 60$\,s and $\tau_{\mathrm{Ch}} = 120$\,s respectively. Additional slow OU noise ($\theta_{\mathrm{slow}} = 0.05$\,s$^{-1}$) accounts for weather-driven variability and constitutes the physical source of the thermal correlation tier: facilities subject to common climate conditions share correlated HVAC and Chiller disturbances independent of their IT workload state. State transitions across the three facilities are independent, so spatial correlation arises from workload overlap statistics rather than from externally imposed phase-locking. The result is the episodic, non-stationary coherence that the DMD method is designed to characterize.

\vspace{-0.5em}
\subsection{RTDS Test System Configuration}
The IEEE 39-bus New England test system is implemented on the RTDS platform. Three converter-interfaced AI data center loads are placed at buses 4, 12, and 15, each with a co-located battery energy storage system at a 1:4 energy-to-power ratio, rated at 200\,MW, 225\,MW, and 250\,MW respectively following \cite{Chaudhary2026SpatialAI, Chaudhary2025DataCenterStability}. \rev{Each data center occupies one bus, so the three facilities map one-to-one to buses 4, 12, and 15, and their three unique bus pairs form the three-component correlation state.} \rev{Buses 4, 12, and 15 are pure load buses on distinct inter-area corridors separated by multiple line impedances, chosen so that any observed inter-bus coherence reflects workload-driven dynamics propagated through the network rather than local electrical proximity~\cite{Chaudhary2026SpatialAI}.} \rev{A co-located BESS is a standard data center design preference, included to provide about 15 minutes of backup at rated power.} Synthetic profiles from Section~IV-A serve as time-varying active and reactive power setpoints. Bus voltage magnitudes, angle deviations, and frequency recordings are the primary outputs alongside the active power at each data center bus. Total vector error in the measurement chain is maintained below 1\% per IEEE~C37.118.1 \cite{ieee_c37_118_1_2011}.
\enlargethispage{4\baselineskip}

%% file: Contents/Results.tex
\subsection{Time-Domain System Response}
Fig.~\ref{fig:time_series_load_volt_freq}(a) illustrates the active and reactive power profiles at the three data center buses.
Active power peaks occur nearly simultaneously across all locations. Reactive power follows with slight phase offsets attributable to independent per-converter voltage regulators.
Fig.~\ref{fig:time_series_load_volt_freq}(b \& c) shows the corresponding RTDS bus voltage magnitudes and system frequency over the same record.
All three buses maintain voltages within $\pm$5\% of nominal ($1.04$--$1.07$~p.u.), and system frequency remains within $[59.93,\,60.07]$~Hz ($\Delta f_{\text{std}} = 0.00403$~Hz).
The absence of large steady-state deviations confirms that converter-interfaced data center loads do not destabilize the IEEE 39-bus system under nominal operating conditions. The stability risk lies instead in the \textit{spatially correlated transient} behavior identified in the following analysis.
\enlargethispage{2\baselineskip}



\subsection{Correlation State Vector and Spatial Index}
\label{subsec:corr_state}
The spatial concentration index $\rho_s(t)$ exceeds the operational threshold of 0.5 in 98.7\% of all windows, with a time-mean of 0.7375 and a maximum of 0.9851.
\rev{The index $\rho_s = \lambda_1 / \sum_k \lambda_k$ is the fraction of total fluctuation variance carried by the dominant spatial mode. For $N_b = 3$ independent loads this fraction equals $1/N_b = 0.333$, which defines the independence baseline, and full single-mode coherence drives it to unity. The operational threshold of 0.5 marks the level at which the dominant mode captures the majority of total variance, well above the independence baseline.}
This persistent elevation, concurrent with near-zero time-averaged pairwise correlations, is the defining signature of non-stationary episodic coherence, where the three-bus ensemble is collectively correlated in direction even when individual pairwise coefficients oscillate in sign.

\begin{figure}[!htbp]
  \centering
  \vspace{-0.5em}
  \includegraphics[width=0.9\columnwidth]{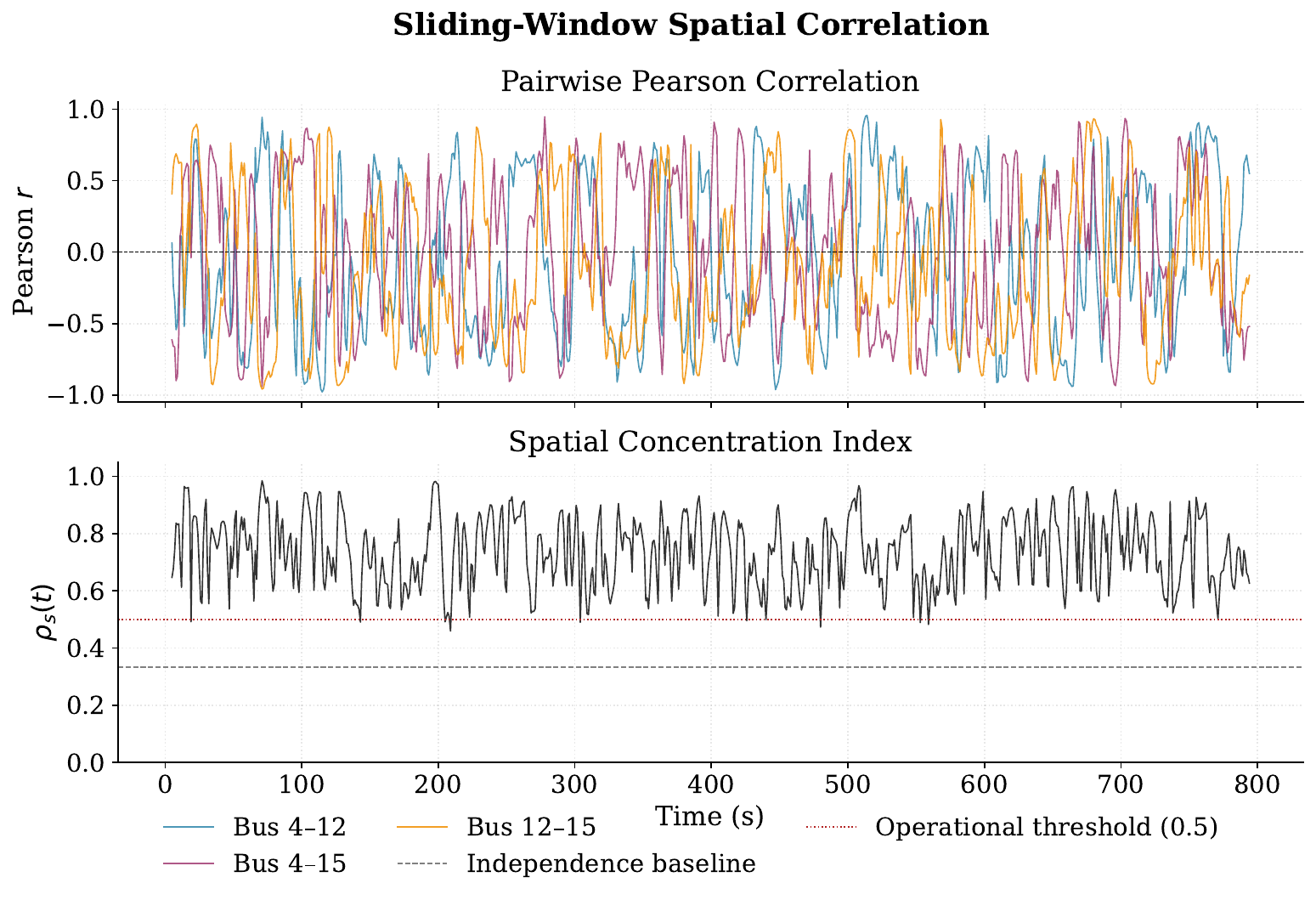}
  \vspace{-1.em}
  \caption{Sliding-window pairwise Pearson correlation \& spatial concentration index $\rho_s(t)$}
  \label{fig:sliding_corr}
  \vspace{-1.25em}
\end{figure}

Fig.~\ref{fig:sliding_corr} presents the sliding-window pairwise Pearson coefficients and $\rho_s(t)$ over the full record.
Each pair traverses nearly the full correlation range $[-1,+1]$, yet time-averaged means are near zero for all pairs. Per-pair standard deviations of about 0.51 confirm that a scalar time-average would report near-independence while the signal exhibits strong intermittent coupling in both directions.

\enlargethispage{6\baselineskip}

\subsection{Global DMD: Eigenvalue Constellation}
\label{subsec:global_dmd}
\begin{figure}[H]
  \centering
  \includegraphics[width=\columnwidth]{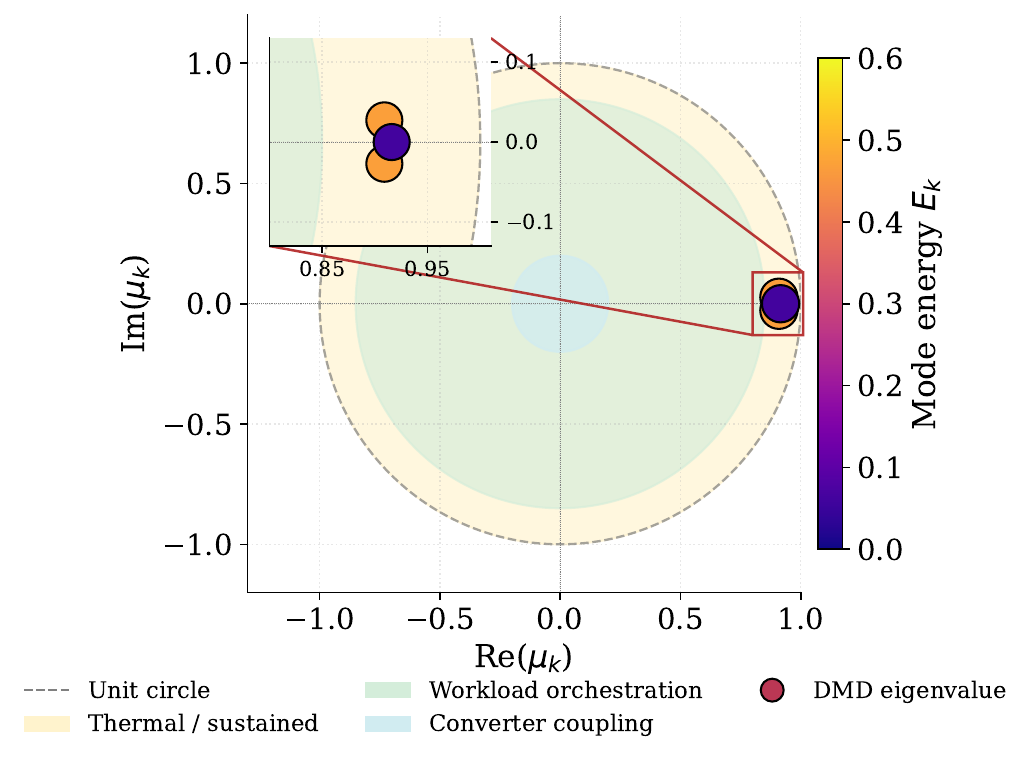}
  \vspace{-1.75em}
  \caption{\rev{Global DMD eigenvalue constellation on the complex plane, with a thermal-band zoom inset. All three modes lie in the slow/thermal band just inside the unit circle, so the time-averaged model resolves only thermal coherence.}}
  \label{fig:global_dmd}
\end{figure}

\rev{The modal results read directly at the level of the inter-bus correlations. An eigenvalue magnitude states whether the inter-bus correlation is growing, steady, or decaying, its frequency states the timescale of the driving mechanism, and its mode shape states which bus pairs participate.}
Fig.~\ref{fig:global_dmd} presents the global DMD results.
The three singular values are $\sigma_1 = 15.64$, $\sigma_2 = 14.35$, and $\sigma_3 = 13.14$, with an overall gap ratio $\sigma_1/\sigma_r = 1.19$. The absence of a pronounced spectral gap is itself a primary finding: flat singular values arise when correlation dynamics span the full available subspace across distinct regimes rather than concentrating in a fixed low-dimensional subspace.
The rank-3 reconstruction error $\varepsilon_r = 0.9982$ quantifies this non-stationarity: a single stationary rank-3 model cannot approximate a signal whose correlation structure alternates between coherence episodes and decorrelated transients. Accordingly, the global eigenvalues are interpreted as time-averaged descriptors rather than a predictive dynamical model.

\begin{figure}[!htbp]
    \centering
    \vspace{-1em}
    \includegraphics[width=1.0\columnwidth]{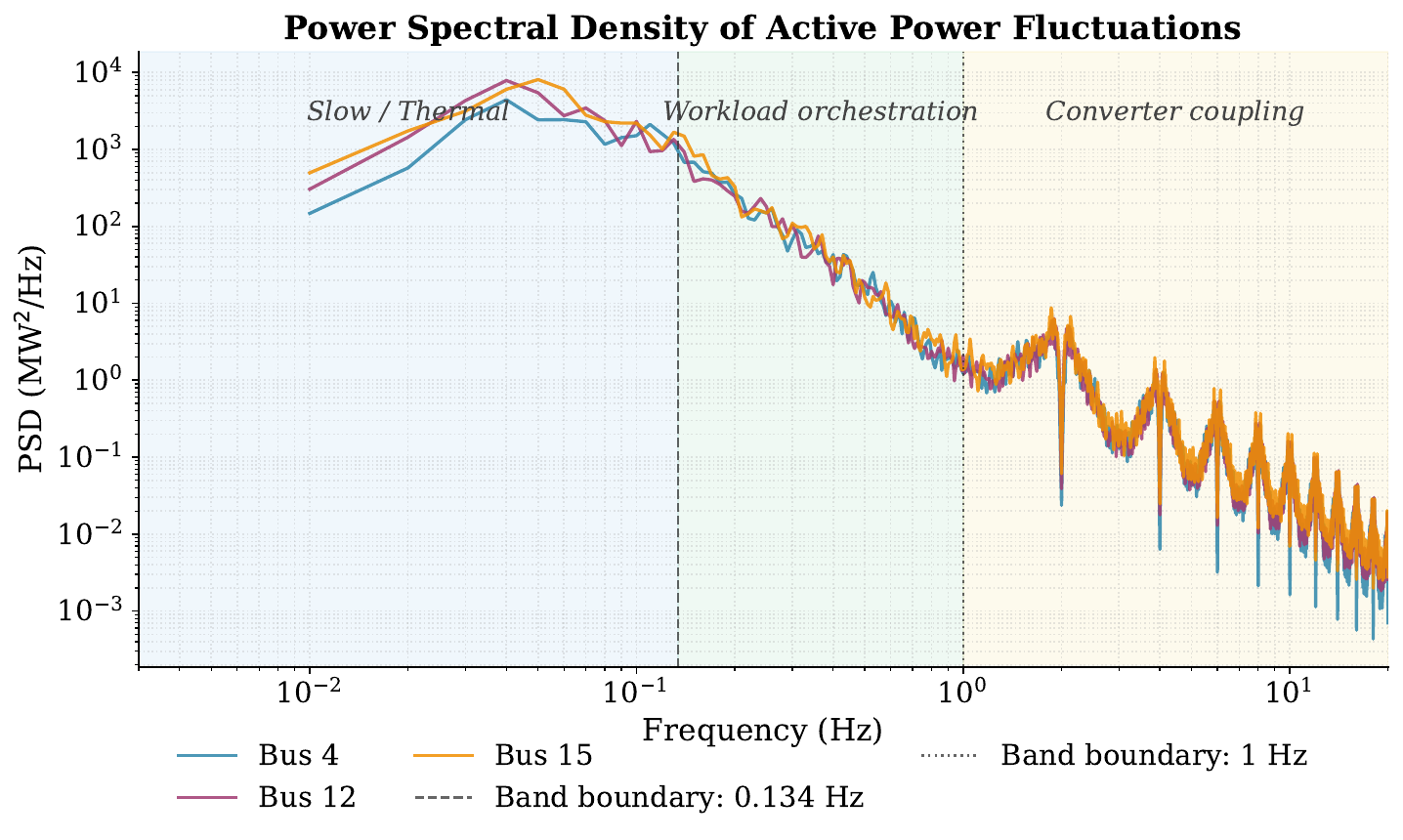}
      \vspace{-1.75em}
    \caption{Power Spectral Density of Power Fluctuations}
    \label{fig:power_spectral_density}\vspace{-1em}
\end{figure}

Fig. \ref{fig:power_spectral_density} shows the power spectral density of the power fluctuations. All three global modes in the time-averaged spectrum fall in the Slow/Thermal band ($0.003$--$0.134$~Hz), and Table~\ref{tab:dmd_modes} summarises their properties.
Modes~1 and 2 form a conjugate pair at $f = \pm 0.0047$~Hz that carries 93.6\% of the total mode energy, and mode~3 is a near-DC mode with the remaining 6.3\%. All eigenvalues reside well inside the unit circle, which reflects net correlation decay averaged over the full record. The quasi-uniform mode shape magnitudes across electrically distinct bus pairs (max/min ratio 1.16--1.38) exclude impedance-mediated converter coupling as the primary driver. The uniform structure instead points to a facility-level thermal mechanism that acts simultaneously on all three buses.

\enlargethispage{5\baselineskip}
\begin{table}[!htbp]
  \centering
  \vspace{-1.75em}
  \caption{Global DMD Mode Summary}
  \label{tab:dmd_modes}
  \vspace{-0.25em}
  \begin{tabular}{cccccc}
    \toprule
    \textbf{Mode} & $f$\textbf{(Hz)} & $|\mu_k|$ & $\sigma_k$\textbf{(s$^{-1}$)} & $E_k$ & \textbf{Band} \\
    \midrule
    1 & $+0.0047$ & $0.909$ & $-0.095$ & $0.468$ & Slow/Thermal \\
    2 & $-0.0047$ & $0.909$ & $-0.095$ & $0.468$ & Slow/Thermal \\
    3 & $\phantom{+}0$ & $0.916$ & $-0.088$ & $0.063$ & Slow/Thermal \\
    \midrule
    \multicolumn{6}{c}{\small Mode shape $|\phi_k|$: Bus 4--12 \quad Bus 4--15 \quad Bus 12--15} \\
    \midrule
    1 \& 2 & $0.445$ & $0.368$ & $0.508$ & \multicolumn{2}{c}{max/min = 1.38} \\
    3       & $0.561$ & $0.538$ & $0.485$ & \multicolumn{2}{c}{max/min = 1.16} \\
    \bottomrule
  \end{tabular}
  \vspace{-1em}
\end{table}

\rev{\textit{Limitations of Global DMD.} Notably, the rank-3 model does not produce eigenvalues in either the workload orchestration band or the converter control band. Two mechanisms explain this absence.}

\subsubsection{Converter Control Coupling}
\rev{Fast converter dynamics are effectively decoupled from the slower thermal behavior. The correlation state is sampled once per second, so an oscillation faster than one half-cycle per second cannot be represented and folds toward the origin rather than the unit circle. Converter control at 1--26.1~Hz lies far above this limit, so the one-second interval isolates the slow thermal correlation dynamics from converter switching.}

\subsubsection{Workload Orchestration}
The three AI load profiles are driven by \textit{independent} semi-Markov state machines with no inter-facility coordination.
Spatial correlation arises only when all three facilities happen to enter the same state simultaneously, a brief (2--30~s) and infrequent coincidence across the 799.5~s record. These transient excursions produce DMD eigenvalues deep inside the unit circle ($|\mu_k| \ll 1$, low energy $E_k$) with no sustained repeating frequency. The episodic, aperiodic nature of workload correlation is precisely what the \textit{sliding-window portrait} resolves through the $|\mu_k^{(n)}|>1$ criterion. Both workload and converter coupling remain consequential for grid stability and are recovered by the sliding-window portrait and the cross-validation that follow.
\enlargethispage{5\baselineskip}

\subsection{Sliding-Window DMD Portrait}
\label{subsec:sliding_dmd}
\begin{figure}[!htbp]
  \centering
  \vspace{-0.5em}
  \includegraphics[width=0.95\columnwidth]{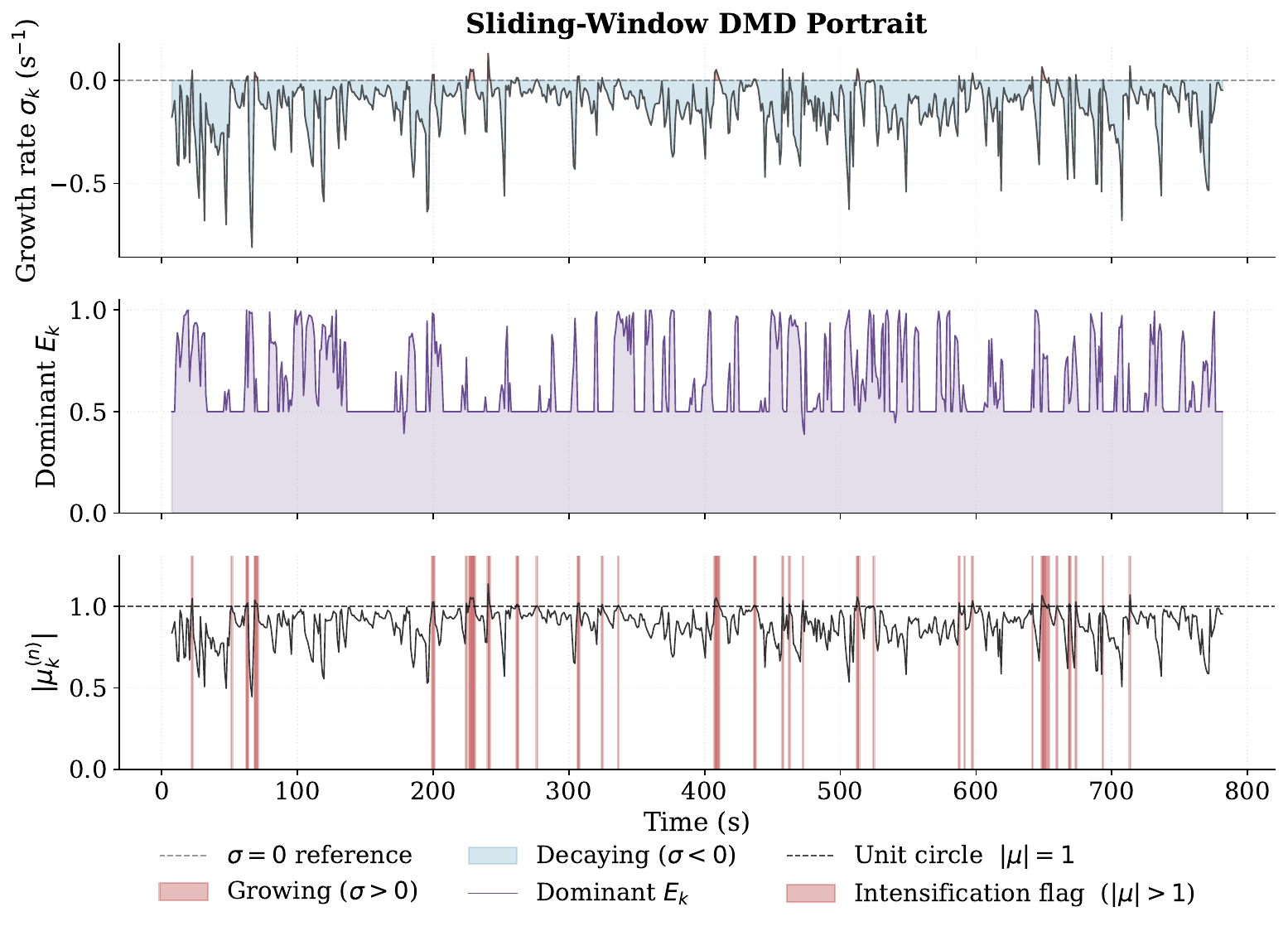}
  \vspace{-1em}
  \caption{Sliding-window DMD portrait. Top: growth rate $\sigma_k^{(n)}$. Middle: dominant mode energy $E_k^{(n)}$. Bottom: $|\mu_k^{(n)}|$ with unit-circle threshold}
  \label{fig:portrait}
\end{figure}

Fig.~\ref{fig:portrait} presents the sliding-window DMD portrait from 775 windows of $T_w^{\mathrm{DMD}} = 15$~s advanced at $\delta t = 1$~s. In 51 of 775 windows (6.6\%), $|\mu_k^{(n)}| > 1$, an intensification flag; flagged windows cluster at $t \approx 22$~s, $62$--$70$~s, and recurrently near $t = 200,\;225,\;300,\;410,\;450,\;520,\;600,\;650,\;700$~s, consistent with the stochastic workload model in which concurrent Surge-state occupation lasts several seconds before asynchronous state evolution separates the buses. The dominant mode energy $E_k^{(n)}$ exhibits rapid transitions between high-concentration episodes ($E_k\!\to\! 1$) and diffuse episodes ($E_k \approx 0.5$), time-aligned with the intensification flags, which identifies two distinct dynamical regimes: a phase-locked regime where one mode captures nearly all state-vector variance, and a decorrelated regime where dynamics are spread across the available subspace. 

The $|\mu_k^{(n)}| > 1$ criterion constitutes a precursor indicator, detectable before pairwise coefficients reach their peak, suitable for deployment as a streaming operational alert. \rev{Across the 31 intensification episodes, the median lead from the flag onset to the subsequent peak of the dominant pairwise correlation is 4.0~s, with a mean of 4.4~s and a range of 0--11~s, and 74\% of episodes provide at least 2~s of lead. This seconds-scale lead reflects the short duration of the coherence episodes, and longer anticipatory horizons follow from the autoregressive amplitude predictor described in the conclusion.}

\subsection{Cross-Validation Against RTDS Voltage Signals}
\label{subsec:xval}
The cross-validation is performed through magnitude-squared coherence $\gamma^2_{ij}(\omega)$ of voltage deviations at flagged and sparse episodes.
For Bus~4--12, the flagged-episode dominant peak at 0.366~Hz ($\gamma^2 = 0.962$) falls in the workload orchestration band; during sparse episodes the peak shifts to 16.6~Hz ($\gamma^2 = 0.992$) in the converter band. The Slow/Thermal band peak shifts from $\gamma^2 = 0.807$ at 0.061~Hz (flagged) to $\gamma^2 = 0.954$ at 0.092~Hz (sparse), directly cross-validating the load-domain DMD portrait. Bus~12--15 exhibits a topology-dependent reversal: the flagged-episode Slow/Thermal peak ($\gamma^2 = 0.550$ at 0.122~Hz) falls \textit{below} the sparse-episode value ($\gamma^2 = 0.921$ at 0.031~Hz), while both episodes show high converter-band coherence, with the flagged peak at 14.77~Hz ($\gamma^2 = 0.919$) consistent with documented incidents~\cite{mishra2025understanding}. The reversal arises because buses~12 and 15 are connected through higher-impedance network paths, attenuating thermal-coupling coherence during intensification events.

\enlargethispage{5\baselineskip}


Collectively, the intensification flags correspond to episodes of elevated inter-bus voltage coherence at the frequencies the DMD method predicts, which confirms that the modal portrait reflects network-level coupling rather than a construction artifact.

%% file: Contents/conclusion.tex
This paper proposed a DMD-based method for characterizing non-stationary spatial load correlation in AI data center-dominated power systems. The method addresses a gap that time-averaged spectral methods cannot close. By applying DMD to the temporal evolution of the correlation state vector, the method extracts physically interpretable modes and growth indicators without a stationarity assumption, and produces an early-warning signal detectable before pairwise correlations reach their peak. While global DMD attributes the dominant energy to slow thermal dynamics, the sliding-window portrait reveals that workload orchestration contributes intermittent intensification events that global analysis alone would miss. Cross-validation with RTDS bus voltage coherence confirms that both findings reflect genuine network-level coupling. The results establish that the modal growth indicator is a viable operational diagnostic for the episodic inter-bus coherence that characterizes AI data center clusters at transmission scale.

Several directions remain open. An autoregressive model of the dominant mode amplitude can predict spatial concentration index exceedance over a short horizon, which gives operators lead time for reserve pre-positioning and correlation-aware dispatch. An extension with compressed DMD and sparse sensor placement can recover the modal portrait from a limited subset of buses and validate it on larger data center clusters beyond the three-bus configuration considered here.

%% file: Contents/Acknowledgment.tex
The research is supported in part by MSU Research Foundation and in part by the U.S. National Science Foundation under grant 2408615.

This article has been authored by employees of National Technology \& Engineering Solutions of Sandia, LLC under Contract No. DE-NA0003525 with the U.S. Department of Energy (DOE). The employee owns all right, title and interest in and to the article and is solely responsible for its contents. The United States Government retains and the publisher, by accepting the article for publication, acknowledges that the United States Government retains a non-exclusive, paid-up, irrevocable, world-wide license to publish or reproduce the published form of this article or allow others to do so, for United States Government purposes. The DOE will provide public access to these results of federally sponsored research in accordance with the DOE Public Access Plan
\href{https://www.energy.gov/downloads/doe-public-access-plan}{https://www.energy.gov/downloads/doe-public-access-plan}. 